\documentclass[runningheads]{cl2emult}

\usepackage{makeidx}  % allows index generation
\usepackage{graphicx} % standard LaTeX graphics tool
\usepackage{subeqnar} % subnumbers individual equations
\usepackage{multicol} % used for the two-column index
\usepackage{cropmark} % cropmarks for pages without
\usepackage{eso}      % placeholder for figures
\makeindex            % used for the subject index
\begin{document}
\title*{Did the Universe start at Zero Metallicity?}
\toctitle{Did the Universe start at Zero Metallicity?
\protect\newline }
% allows explicit linebreak for the table of content
%
%
\titlerunning{Did the Universe start at Zero Metallicity?}
% allows abbreviation of title, if the full title is too long
% to fit in the running head
%
\author{Karsten Jedamzik}       
%\and Roger Temam\inst{2}
%\and Jeffrey Dean\inst{2}
%\and David Grove\inst{1}
%\and Craig Chambers\inst{2}
%\and Kim~B.~Bruce\inst{2}
%\and Elsa Bertino\inst{1}}
%
\authorrunning{Karsten Jedamzik}
% if there are more than two authors,
% please abbreviate author list for running head
%
%
\institute{Max-Planck-Institut f\"ur Astrophysik, 85740 Garching, Germany}
%\and Universit\'{e} de Paris-Sud,
%     Laboratoire d'Analyse Num\'{e}rique,
%     B\^{a}timent 425,\\
%     F-91405 Orsay Cedex, France}

\maketitle              % typesets the title of the contribution

 \begin{abstract}
Standard Big Bang nucleosynthesis predicts an essentially zero
primordial metallicity. I speculate on possible metal (i.e. nucleon number
$A\geq 12$) production in scenarios of inhomogeneous Big Bang
nucleosynthesis. It is conceivable, though not necessarily probable, 
that some primordial metallicity is synthesized if a small fraction of
all cosmic baryons reside in very high-density regions. Such
conditions could possibly result from the
evaporation of some baryon-number carrying soliton prior to the epoch
of Big Bang nucleosynthesis.
 \end{abstract}

%\section{Fixed-Period Problems: The Sublinear Case}

It is widely accepted that an era of 
Big Bang nucleosynthesis (hereafter; BBN) between cosmic temperatures
$T\sim 1$ MeV and $T\sim 10$ keV is the production site of the bulk of $^4$He 
and $^3$He,
a good fraction of $^7$Li, and essentially all $^2$H 
observed in nebulae and stars in
the present universe. Generally good 
agreement between observationally inferred 
primordial abundances and theoretical predictions for $^2$H, $^4$He, and 
$^7$Li may be
obtained within a standard homogeneous BBN scenario with baryon-to-photon
ratio $\eta$ in the range $2\times 10^{-10} - 6\times
10^{-10}$~\cite{BBN} 
(most likely
towards the upper end of this range), 
though details of the observational determination 
of primordial abundances and the question about consistency within a
standard BBN scenario remain under investigation.
Primordial conditions as envisioned in a standard BBN scenario yield only
production of minute amounts of isotopes with nucleon number
$A>7$. For a baryon-to-photon
ratio of $\eta\approx 4\times 10^{-10}$ synthesis of a $^{12}$C 
mass fraction of only
${\rm X_{^{12}{\rm C}}}\approx 5\times 10^{-15}$ results. 
There is also production of trace amounts
of $^6$Li, $^9$Be, and $^{11}$B, with typical mass fractions $\sim 10^{-16} -
10^{-13}$. 
In contrast to stellar nucleosynthesis, the three-body
triple alpha process is not operative in standard BBN, 
mainly due to the small densities ($\sim 10^{-5}$g/cm$^3$ at $T\approx 100$
keV) 
and short expansion time ($\sim 100$ s). A potentially different way of bridging 
between light and \lq\lq heavy\rq\rq\ ($A\geq 12$) 
elements across the mass eight
gap is via the reaction sequence $^7$Be $\to$ $^{11}$C $\to$ 
$^{11}$B $\to$ $^{12}$C. This reaction
chain is, nevertheless,
ineffective in a standard BBN since build-up of $^7$Be occurs late at 
$T\sim 50-100$ keV where Coulomb barrier effects 
prevent further processing of this isotope
into heavier ones.

One may pose the following question: {\it Which alternative BBN scenarios
would give substantial initial metal production, without violating
observational constraints on the light elements?} Here \lq\lq metal\rq\rq 
refers to isotopes with nucleon number $A\geq 12$ and by the term \lq\lq
substantial\rq\rq\ it is meant that production would result in abundance which
could either be directly observable in the not-to-distant future within the 
atmospheres of metal-poor stars, 
or would be sufficient for the operation of the CNO
cycle within the first stars (${\rm X_{CNO}}>10^{-10}$). The only scenario for 
pre-stellar production of metals known to the author 
is a BBN scenario characterized by an inhomogeneous baryon distribution.
Such inhomogeneities in the baryon-to-photon ratio may result from the
out-of-equilibrium dynamics of the early universe prior to the BBN era, 
as possibly during a
first-order QCD phase transition ($T\approx 100$ MeV) or an era of 
inhomogeneous
electroweak baryogenesis ($T\approx 100$ GeV). Another possibility may be the
creation of baryon \lq\lq lumps\rq\rq via the evaporation of baryon
number carrying solitons formed in the early universe, 
such as strange quark matter nuggets or B-balls. Given the wide variety of
possibilities and the large intrinsic uncertainties of 
scenarios for the generation
of baryon number fluctuations in the early universe, it is probably best to initially
regard the question about primordial abundance yields and possible metallicity
production in 
inhomogeneous BBN scenarios disconnected from the speculative
mechanism for
the creation of baryon number fluctuations. 

In the past, there have been a number of investigations of the evolution of
pre-existing $\eta$ fluctuations from high cosmic temperatures ($T\approx
100$ GeV)
through the epoch of BBN~\cite{EVO}, and of the resulting BBN
abundance yields~\cite{IBBN}.
Baryon number fluctuations only impact BBN yields if they survive
complete dissipation by neutron diffusion
before the epoch of weak freeze-out at $T\approx 1$ MeV. 
This is the case for fluctuations containing baryon number in excess
of $N_b\, ^>_{\sim}\, 10^{35}(\eta_h/10^{-4})^{-1/2}$, 
where $\eta_h$ is the  baryon-to-photon ratio within the fluctuation,
corresponding to a baryon mass of $M_b\, ^>_{\sim}\, 10^{-22}M_{\odot}
(\eta_h/10^{-4})^{-1/2}$. 
For reference, the baryon mass
contained within the QCD- and electroweak horizon are 
$\sim 10^{-8}M_{\odot}$ and $\sim 10^{-18}M_{\odot}$, respectively.
The evolution of fluctuations after the epoch of weak freeze-out,
for a large part of parameter space, is 
characterized by differential neutron-proton diffusion resulting in 
high-density, proton-rich
and low-density, neutron-rich regions. 
It had been suggested that such scenarios may not only
result in consistency between
observationally inferred and theoretically predicted primordial abundances
for much larger horizon-average baryon-to-photon ratios than $\eta\approx
4\times 10^{-10}$, but also lead to a possible
efficient r-process nucleosynthesis~\cite{Apple} 
in the neutron-rich regions. When all
dissipative and hydrodynamic processes during inhomogeneous BBN are 
treated properly, one finds that, except for a fairly small parameter
space, the
average $\eta$ in inhomogeneous BBN scenarios may not be larger than in
standard BBN due to overabundant $^7$Li and/or $^4$He 
synthesis~\cite{Result}. A detailed
investigation of possible r-process nucleosynthesis in the neutron-rich regions
shows that an r-process is utterly ineffective in inhomogeneous 
BBN~\cite{r-process}. 
This is mostly due to r-process yields being a sensitive function of the local 
baryon-to-photon ratio
in the neutron-rich region, which generically is too low in inhomogeneous BBN.

Nevertheless, BBN metal synthesis via $p$- and $\alpha$-burning reactions
of cosmologically interesting magnitude
may occur if a small fraction, $f_b$, of the cosmic baryons reside in regions 
with very high baryon-to-photon ratio, 
$\eta_h\geq 10^{-4}$~\cite{metals}. Whereas
inhomogeneous BBN scenarios are typically plagued by overproduction of
$^7$Li,
the comparatively early production of $^7$Be 
(which would decay into $^7$Li) in regions with
$\eta_h\geq 10^{-4}$ allows for a further processing of $^7$Be 
into heavier isotopes.
The net nucleosynthesis of such regions are a $^4$He mass fraction $Y_p\approx
0.36$, some metals, and virtually no $^2$H, $^3$He, and $^7$Li. 
A baryon-to-photon ratio $\eta\approx 10^{-4}$ is interesting as it presents
the asymptotic value attained after partial dissipation 
of initially very high-density regions
$\eta_i\gg 10^{-4}$ by the action of 
neutrino heat conduction at temperature $T\gg
1$ MeV, for a wide part of parameter space. 
Baryon \lq\lq lumps\rq\rq\ with initial baryon-to-photon ratio
$\eta_i$ and baryonic mass less than $M_b\leq 10^{-14}M_{\odot}
\eta_i$ will have evolved to an asymptotic $\eta_f\approx 10^{-4}$
by the time of weak freeze-out.  They are further almost unaffected by neutron
diffusion during BBN provided their mass exceeds $M_b \, ^>_{\sim}\, 
10^{-19}M_{\odot}$. 
In this case horizon-averaged BBN abundance yields may be determined by
an appropriate average of the yields of two homogeneous universes, one at
$\eta_h\approx 10^{-4}$ and one at $\eta_l$, close to the observationally 
inferred value for a standard BBN scenario. The resultant cosmic 
average metallicity in such an inhomogeneous scenario
may be approximated by~\cite{metals}
\begin{equation}
\nonumber
[{\rm Z}]\sim -6.5 + {\rm log}_{10}(f_b/10^{-2}) + 2\,
{\rm log}_{10}(\eta_h/10^{-4})\, ,
\end{equation}
where [Z] denotes the logarithm of ${\rm X}_{A\geq 12}$ 
relative to the solar value
(${\rm X}_{\odot}\approx 2\times 10^{-2}$). 
The most stringent constraint on
the fraction of baryons residing in high $\eta$ regions comes from a possible
overproduction of $^4$He, with a change $\Delta Y_p\approx 0.12f_b$ relative to
a homogeneous BBN at $\eta_l$, allowing for $f_b\sim 10^{-2}-3\times 10^{-2}$,
and [Z]$\sim -6$, possibly even larger for $\eta_h > 10^{-4}$.
Given $\eta_h\approx 10^{-4}$, a nucleosynthetic signature is given
by [O/C] $\sim 1.3$
and [C/$Z_{A\geq 28}]\sim -1.5$. 
For $\eta_h\gg 10^{-4}$, synthesis of mostly iron-group elements results.

In conclusion, it is possible, though not necessarily probable, that the universe
\lq\lq started\rq\rq\ with some initial metallicity if a small fraction of
the cosmic baryon number existed in high-density regions at the epoch of BBN,
This metallicity may be large enough to have the first stars operate 
on the CNO-cycle, even though it probably would fall
below the metallicities observed in the currently 
most metal-poor stars ([Z] $\sim -5$) known.

%INDEX%%%%%%%%%%%%%%%%%%%%%%%%%%%%%%%%%%%%%%%%%%%%%%%%%%%%%%%%%%%%%%%
\clearpage
\addcontentsline{toc}{section}{Index}
\flushbottom
\printindex
%%%%%%%%%%%%%%%%%%%%%%%%%%%%%%%%%%%%%%%%%%%%%%%%%%%%%%%%%%%%%%%%%%%%%

\end{document}